ORIGINAL ARTICLE

# Experimental Pathways for Detecting Double Superionicity in Planetary Ices

Kyla de Villa[1] | Felipe González-Cataldo[1] | Burkhard Militzer[1,2]

[1]Department of Earth and Planetary Science, University of California, Berkeley, California, USA

[2]Department of Astronomy, University of California, Berkeley, California, USA

**Correspondence**
Corresponding author Kyla de Villa.
Email: kyla.devilla@berkeley.edu

**Abstract**

The ice giant planets Uranus and Neptune are assumed to contain large amounts of planetary ices such as water, methane, and ammonia. The properties of mixtures of such ices at the extreme pressures and temperatures of planetary interiors are not yet well understood. *Ab initio* computer simulations predicted that a number of ices exhibit a hydrogen superionic state and a doubly superionic state [DOI 10.1038/s41467-023-42958-0]. Since the latter state has not yet been generated with experiments, we outline here two possible pathways for reaching and detecting such a state with dynamic compression experiments. We suggest X-ray diffraction as the principal tool for detecting when the material becomes doubly superionic and the sublattice of one of the heavy nuclei melts. That would require a temperature of ∼3500 K and pressures greater than ∼200 GPa for $H_3NO_4$, which we use as an example material here. Such conditions can be reached with experiments that employ an initial shock that is followed by a ramp compression wave. Alternatively, one may use triple-shock compression because a single shock does not yield sufficiently high densities.

**KEYWORDS**

superionic, high pressure, planetary ices, Uranus, Neptune, ice giant planets, X-ray diffraction, shock Hugoniot curve, density functional theory, molecular dynamics

## 1 | INTRODUCTION

The ice giants Uranus and Neptune orbit our sun at distances of 19 and 30 astronomical units, respectively, making them difficult to visit by spacecraft. Consequently, they are relatively unexplored, with open questions remaining regarding their interior structures, thermal profiles, dynamic processes, and even composition. Following the rise of exoplanet detection methods in the last 30 years, such questions now also extend to super-Earth and sub-Neptune sized exoplanets, which dominate currently discovered exoplanets[1]. Although the exact proportion of rock to ice in these planets is uncertain, it is typically assumed that Uranus and Neptune are composed of at least 50% planetary ices - mixtures of water, methane, and ammonia accreted in large volumes by these planets during their formation[2]; but the compositional space possible for icy exoplanets is far broader.

At the present time, the properties of mixtures of planetary ices are relatively unexplored at extreme conditions, with most studies focused on pure materials, binary mixtures, or specific stoichiometric mixtures such as "synthetic Uranus,"[3] with a H:C:N:O ratio of 27:7:4:1, an approximate representative of proto-solar elemental ratios[4,5,6].

Recently, crystal structure prediction methods have been applied to infer stable crystalline phases of H-C-N-O materials which may form from $H_2O$, $NH_3$, and $CH_4$ at the extreme pressures of planetary interiors[7,8]. An investigation of the high-temperature behaviors of the hydrogen bearing materials predicted through these structure searches revealed the prevalence of the hydrogen superionic phase of planetary ices at planetary interior conditions[9,10,11,12]. This phase is characterized by the rapid diffusion of H ions through a stable lattice of the heavier ions in the material, and has been observed computationally and indirectly through experiments for the H-C-N-O chemical space for water, ammonia, ammonia hydrides, and synthetic Uranus. Computational work has revealed a second superionic phase in many H-C-N-O materials: double superionicity, in which a second element (the lightest of the heavy species C, N, and O) also diffuses superionically, simultaneously with the superionic H ions[12]. Doubly superionic behavior of H-C-O materials was also recently reported, at pressures of 100-500 GPa with simultaneously diffusing H and C atoms[13].

Ion motion in the doubly superionic phase is demonstrated in Fig. 1 for $H_3NO_4$ in the $P2_12_12_1$ phase, which shows the oxygen ions limited to vibrating in their potential wells, while the N and H ions diffuse widely throughout the cell. Figure 2 also illustrates the doubly superionic motion of H and N in $H_3NO_4$-$P2_12_12_1$, by showing the distinct diffusion pathways that each ion type adopts as ions travel through the cell. H and N ion pathways rarely overlap, as H ions primarily orbit O ions at short distances, whereas N ions maximize their distance from O ions.

In their seminal paper, Cavazzoni et al.[9] reported results from density functional molecular dynamics simulations that predicted water and ammonia become superionic at conditions of high pressure and temperature in the interiors of Uranus and Neptune. Superionic water may contribute to the generation of their magnetic fields because the hydrogen nuclei carry an electrical charge so that their diffusion contributes to the material's electrical conductivity. Earlier works had relied on more approximate theoretical methods to predict that water ice exhibits a superionic phase, by applying defect theory in 1985[14], and using classical interatomic potentials in 1988[15]. Indirect experimental confirmation of the existence of superionic phases of planetary ices has only been achieved in recent years. Raman spectroscopy has been employed for both water and ammonia to show the disordering of O-H and N-H bonds, interpreted to signify the onset of proton diffusion[16,17]. Ionic conductivity, combined with compressibility measurements derived from X-ray diffraction (XRD) have been used to argue the observation of the onset of the superionic phase[18]. Similarly, the superionic phase of water has been experimentally verified by showing a difference between electronic conductivity derived from absorption and reflectivity measurements, and total conductivity derived from impedance measurements and density functional theory calculations, which can only be explained by the high ionic conductivity of the superionic phase[11]. Most recently, in situ XRD measurements of shock compressed water have shown the formation of body-centered cubic (bcc) and face-centered cubic (fcc) oxygen sublattices at the temperature and pressure conditions associated with superionic water, experimentally confirming the high-pressure stability of this phase[19].

In this work, we propose a pathway for the experimental detection of the doubly superionic phase of the planetary ice $H_3NO_4$ with symmetry group $P2_12_12_1$ by presenting computed Hugoniot curves and X-ray diffraction patterns based on first principles simulations of shock conditions. We hope that by applying similar experimental approaches as those applied to characterize the properties of superionic water at extreme temperatures and pressures, this novel phase can be observed experimentally for the first time.

## 2 | METHODS

### 2.1 | Molecular Dynamics Simulations

We considered all H-bearing structures predicted to be stable by the H-C-N-O quaternary structure searches executed by Naumova et al.[7] and Conway et al.[8]. Simulations were performed using density functional molecular dynamics (DFT-MD) as implemented in the Vienna ab initio simulation package (VASP)[20].

288 atom cells of $H_3NO_4$-$P2_12_12_1$ were geometrically optimized to desired pressures of 150-550 GPa, and were then simulated in the NVT ensemble equilibrated with the Nose-Hoover thermostat[21,22], using a 0.5 fs timestep. $H_3NO_4$-$P2_12_12_1$ was heated at fixed density from 0 K up to temperatures beyond the melting points in 500 K increments using the heat-until-it-melts approach[23], and thus the melting line shown here represents just an upper bound to the melting temperature.

144 atom cells of a 1:1 liquid $H_2O$ and $HNO_3$ mixture were prepared using the Packmol software[24], followed by geometry optimization at pressures from 1 atm to 100 GPa. These cells were similarly heated up to 2000 K at fixed density with NVT simulations performed until energy equilibration was reached. Upon sufficient pressurization the $H_2O$-$HNO_3$ mixtures exhibited amorphous solid and amorphous superionic phases. Further work is required to constrain the low pressure boundaries between the solid, superionic, doubly superionic, and liquid phases of this mixture.

Electronic structure calculations were completed using the Perdew, Burke, and Ernzerhof (PBE) exchange-correlation functional[25] and a pseudopotential of the projector-augmented wave (PAW)[26] type using a plane-wave cutoff of 1100 eV with core radii of 0.8, 1.5, and 1.1 Å for hydrogen, nitrogen, and oxygen, respectively. The Brillouin zone was sampled with the $\Gamma$-point only which is sufficient for the supercells used in our DFT-MD simulations. Electronic states were populated according to a Fermi-Dirac distribution using the Mermin functional[27] to incorporate the excited electronic states at finite temperature within the Kohn-Sham formalism[28]. These parameters were tested for convergence within 1% for energy, pressure, and components of the stress tensor.

### 2.2 | Shock Hugoniot Calculations

Shock wave experiments are the experimental method of choice when materials at megabar pressures and elevated temperatures need to be studied[29]. The resulting shock Hugoniot curve can be calculated theoretically as long as the equation of state of relevant materials is known with sufficient precision[30,31,32]. At higher temperatures beyond $\sim 2\times 10^5$ K, where nonideal mixing effects have shown to be small[33], one may derive the Hugoniot curve from a linear combination of the EOSs of oxygen[34], nitrogen[35], and hydrogen under the ideal mixing approximation[36,37,38,39,40]. Here we employ DFT-MD simulations to compute the equation of state (EOS) of $H_3NO_4$ directly.

The principal Hugoniot curve was determined by solving for energy, pressure, volume, and temperature conditions that satisfy the Rankine-Hugoniot relationship for shock compression[41]:

$$(E - E_0) + \frac{1}{2}(P + P_0) \times (V - V_0) = 0.$$

Here $E_0$, $P_0$, and $V_0$ represent the initial state of the sample being compressed, which we assume to be a homogeneous liquid $H_2O$-$HNO_3$ mixture at room temperature and pressure. $P_0$ was taken to be 1 atm. As $H_2O$ and $HNO_3$ are liquids at room temperature and pressure, $V_0$ was calculated by assuming a linear mixing of their volumes. $E_0$ was obtained by constructing a cell of 18 $H_2O$ + 18 $HNO_3$ molecules for a total of 144 atoms using the Packmol software[24], followed by geometry optimization.

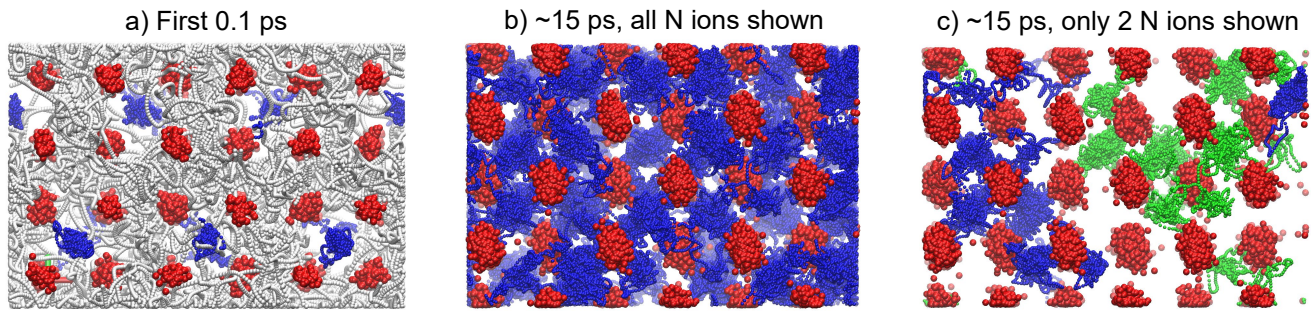

**FIGURE 1** Trajectory visualizations for doubly superionic $H_3NO_4$-$P2_12_12_1$ at 4500 K with a density of 6.033 g/cm$^3$. Individual spheres for each species (O in red, N in blue, and H in white) represent the positions of ions at different time steps, illustrating their motion throughout the simulation. (a) First 0.1 ps of the simulation, demonstrating initial lattice sites of nitrogen and oxygen ions together with H diffusion. Even on such a short timescale, H ions diffuse rapidly. (b) ∼15 ps trajectory showing rapid and widespread N diffusion, whereas oxygen ions vibrate around their potential wells, remaining confined to their lattice sites without diffusing. (c) Same as (b), now with the trajectories of only 2 N ions shown, with one in blue and one in green. For (b) and (c), H trajectories are hidden for ease of visualization of the N trajectories.

## 3 | RESULTS AND DISCUSSION

### 3.1 | Shock Hugoniot Curves

A number of planetary ices have now been suggested to exhibit a doubly superionic phase at the extreme pressures and temperatures of planetary interiors[12,13]. One such compound, $H_3NO_4$ has been predicted[7,8], to have several stable structures, with a $C2_1$ structure exhibiting double superionicity at a density of 5.644 g/cm$^3$, and a $P2_12_12_1$ structure exhibiting double superionicity across a density and pressure range of at least 4.331-6.203 g/cm$^3$ and 150-600 GPa (see Figure 3). The wide pressure range over which $H_3NO_4$-$P2_12_12_1$ exhibits double superionicity makes this material a promising candidate for experimental studies. Moreover, this stoichiometry can be achieved using 1:1 molar ratios of liquid $H_2O$ and $HNO_3$ (nitric acid), which are miscible liquids at room temperature and pressure[42].

With this motivation, we propose two possible shock compression pathways for reaching and detecting the doubly superionic phase of this material, shown in Figures 3 and 4. As these figures show, the principal Hugoniot curve for $H_3NO_4$ does not reach sufficient density to sample the doubly superionic phase. However, such conditions could be reached by employing a single shock followed by ramp compression. Figure 3 shows that ramp compression starting at different conditions along the principal Hugoniot curve would allow each phase of $H_3NO_4$ to be reached at high pressure.

Alternatively, double or triple-shock experiments may be performed to reach the doubly superionic phase of $H_3NO_4$ (see Fig. 4). From the principal Hugoniot, we propose a second shock starting from $T$∼750 K and $\rho$∼2.396 g/cm$^3$, and a third shock starting from $T$∼1500 K and $\rho$∼3.486 g/cm$^3$. The initial conditions for a 1:1 liquid mixture of $H_2O$ and $HNO_3$ and the requisite conditions for the double and triple-shocks are detailed in Table 1. These conditions can be easily achieved by laser platforms such as NIF[43] and the OMEGA Laser Facility[44]. While the initial state of the $H_3NO_4$ is liquid, both ramp compression and the secondary or tertiary shock will pressurize the material into a solid state.

### 3.2 | X-Ray Diffraction Spectra

X-ray diffraction (XRD) is the principal method for detecting phase changes at high pressure[45]. Millot *et al.*[11] employed this technique to determine that superionic water ice assumes a face-centered cubic structure at high pressure as was predicted with DFT-MD simulations by Wilson *et al.*[10]. However, Millot *et al.* relied on indirect arguments about the electrical conductivity to argue that a hydrogen-superionic state had been reached because hydrogen atoms scatter X-rays weakly compared to the heavier oxygen nuclei that come with many more localized core electrons to scatter off of.

The prospects of directly detecting the transition to a doubly superionic state with X-ray diffraction measurements are much better because one of the heavier atomic species becomes disordered and one can expect the XRD peaks associated with that species (nitrogen in the following example) to disappear. We computed the XRD spectra shown in Fig. 5 by averaging over MD trajectories of $H_3NO_4$ performed with 288 atom supercell configurations[46] at a density of 6.033 g/cm$^3$, which corresponds to a pressure of 500 GPa at 0 K. One finds that multiple peaks appear very closely together because the orthorhombic unit cell has the dimensions $a$=2.805 Å, $b$=5.582 Å, and $c$=5.700 Å. Because $2a \approx b \approx c$, the peaks with the Miller indices (2 0 0), (0 4 0), and (0 0 4) appear very close together as they represent Bragg planes with very similar $d$ spacings.

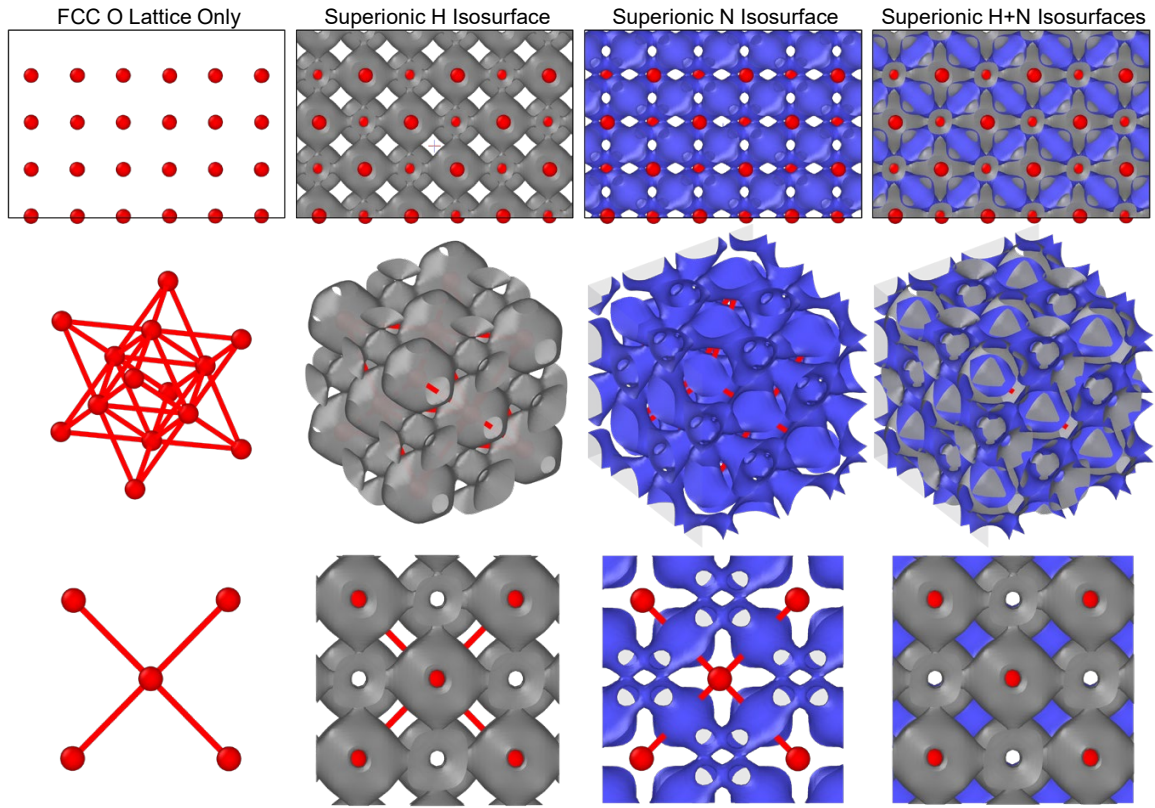

**FIGURE 2** Isosurfaces showing the regions in which superionic ions are most likely to be found in $H_3NO_4$-$P2_12_12_1$ at 4000 K with a density of 6.033 g/cm$^3$. Superionic H is shown in gray, superionic N is shown in blue, with O ions given as red spheres. N and H ions travel through different pathways through the fcc oxygen sublattice, with their pathways rarely crossing. H ions primarily orbit oxygen ions at short distances, whereas N ions maximize their distance from O ions. Top Row: 48 oxygen ion sublattice; Middle Row: fcc oxygen unit cell; Bottom Row: one layer of an fcc oxygen unit cell.

| Stage | T (K) | P (GPa) | $\rho$ (g/cm$^3$) | E (eV/atom) | $U_s$ (km/s) | $U_p$ (km/s) |
|---|---|---|---|---|---|---|
| Initial State | 300 | 0.0001 | 1.358 | −5.496 | – | – |
| Principal Shock | 750 | 18.170 | 2.396 | −5.179 | 5.448 | 2.360 |
| Secondary Shock | 1500 | 86.355 | 3.468 | −4.474 | 9.593 | 2.963 |
| Tertiary Shock | 4000 | 383.992 | 5.325 | −1.989 | 15.687 | 5.472 |

**TABLE 1** Sample conditions at different stages of a triple-shock experiment, corresponding to the suggested compression pathway shown in Fig. 4. The conditions listed for the principal shock are the requisite starting conditions for the secondary shock, and those listed for the secondary shock are the requisite starting conditions for the tertiary shock. The state listed for the tertiary shock is within the predicted boundaries of the doubly superionic phase. These conditions are represented in Figure 4 as cyan hexagons.

We find the XRD spectra for 500 K (solid phase) and 3000 K (hydrogen superionic phase) in Fig. 5 to be very similar. The peak heights are reduced by a modest amount for the hydrogen superionic state at 3000 K. Because hydrogen atoms scatter X-rays only weakly, we attribute the reduction in peak height to the thermal motion of the heavy nuclei due to the Debye-Waller effect[47]. We predict that it will be difficult to detect this reduction directly with XRD measurements as we have seen for $H_2O$ [11]. If one increases the temperatures from 3000 K to 4500 K, the first three peaks in the XRD spectrum in Fig. 5 all disappear because the nitrogen atoms become disordered as the material transitions to a doubly superionic state. In our view, this is the most promising way to detect a doubly superionic state with laboratory experiments. Many other peaks that are associated with an ordered arrangement of oxygen atoms remain present at 4500 K while their peak widths increase because the of thermal motion of the nuclei.

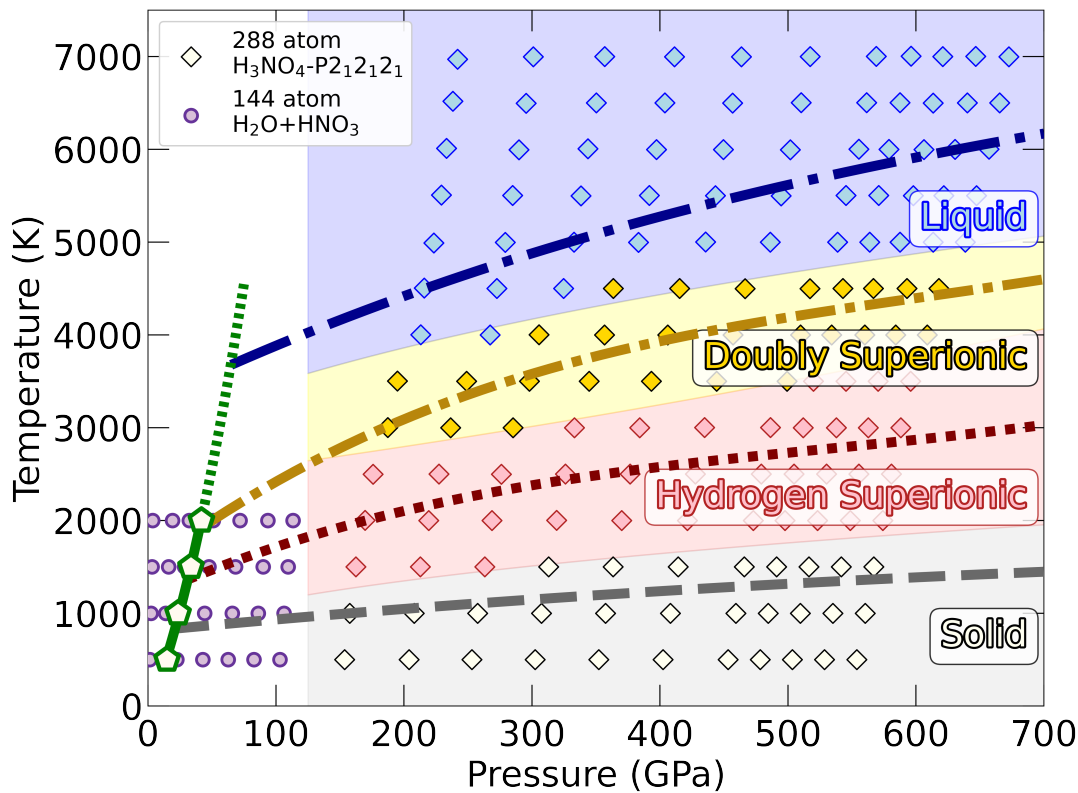

**FIGURE 3** Phase diagram for $H_3NO_4$ showing proposed compression pathways using a single shock followed by ramp compression. The principal Hugoniot curve is shown in green. Isentropes starting at different conditions along the principal Hugoniot curve allow for different phases of $H_3NO_4$ to be reached. Diamonds represent simulations performed starting at 0 K with a $P2_12_12_1$ structure of $H_3NO_4$. Purple circles represent simulations that were started from a 1:1 liquid mixture of $H_2O$ and $HNO_3$.

If the temperature is increased further to 6000 K, the material melts and the computed XRD spectrum exhibits the typical broad peak of a liquid. The peak maximum is centered at the (1 2 2) peak, which is the strongest reflection in all three lower-temperature spectra. This makes this peak essential for detecting that a transition to a solid or superionic phase has occurred during a shock experiment, which we assume started from a mixture of two *liquids* at ambient conditions.

The $H_3NO_4$-$P2_12_12_1$ structure is predicted to be thermodynamically stable at pressures greater than 200 GPa at zero temperature while its stability range at high temperature has not yet been mapped out carefully. Given this uncertainty, one may consider conducting double or triple-shock experiments [48,49,50], ramp compression measurements [51,52,53] or employ diamond cells as a precompression module to increase the initial density of the sample [54,55,56,57,58] to probe deeper in the pressure-temperature field where the $H_3NO_4$-$P2_12_12_1$ structure is strongly favored over other structures.

## 4 | CONCLUSIONS

We presented results from *ab initio* computer simulations of the material $H_3NO_4$ in the $P2_12_12_1$ structure over a broad range of pressures and temperature from 150-700 GPa and 500-7000 K. We discussed pathways for identifying the transition from a solid to hydrogen superionic to a doubly superionic and finally to a liquid state with experimental methods. We showed that XRD measurements are a promising approach to directly detect the transition from a hydrogen superionic to a doubly superionic state. Our calculations show that during this transition, three prominent peaks in the XRD spectrum disappear because the nitrogen atoms become disordered.

We conclude that dynamic compression experiments that employ an initial shock that is followed a ramp wave are the preferred experimental technique to generate the conditions of $T \sim 3500$ K and $P > 200$ GPa where $H_3NO_4$ is predicted to assume a doubly superionic state. Alternatively triple-shock experiments may be employed while a single shock will not yield sufficiently high densities. At the present time, it remains unclear whether double-shock experiments would be able to reach a doubly superionic state because its field of stability in pressure-temperature space has not yet been determined with sufficient accuracy.

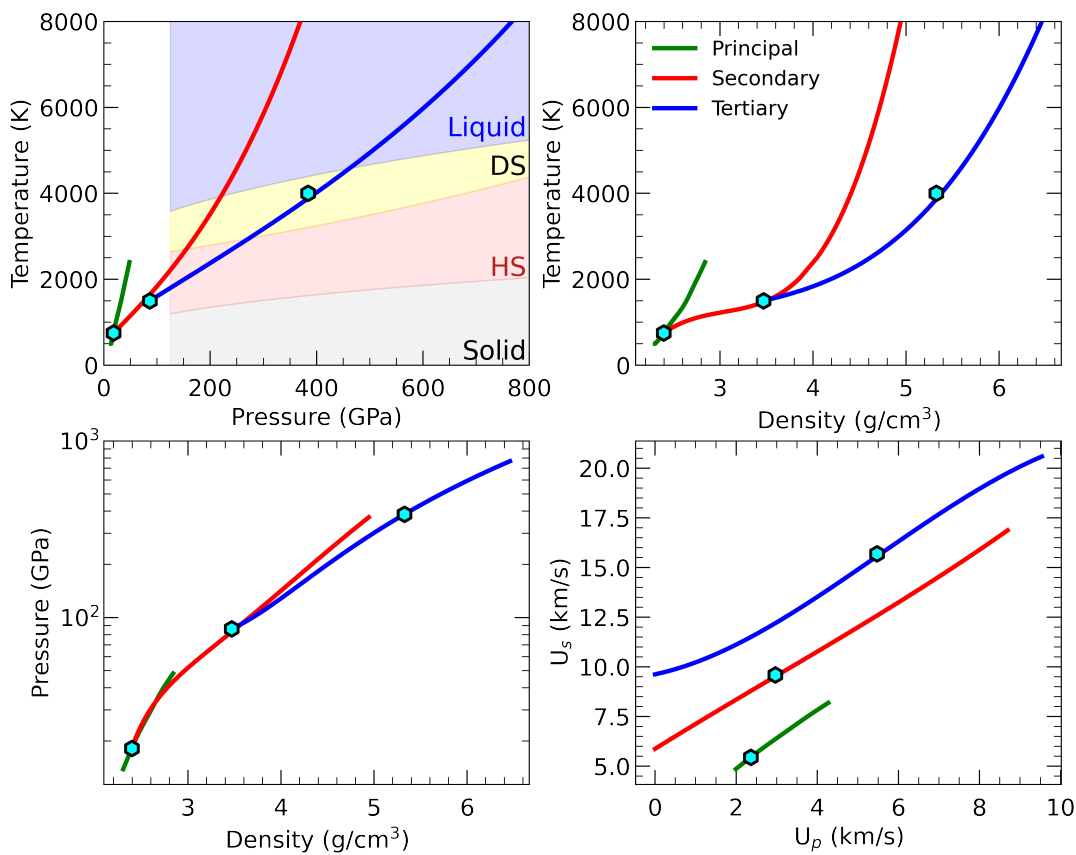

**FIGURE 4** Equations of state of principal, secondary, and tertiary shocks of $H_3NO_4$ starting from a 1:1 mixture of $H_2O$ and $HNO_3$. Top left: pressure/temperature phase diagram showing different shock pathways through doubly superionic phase. Top right: density/temperature phase diagram. Bottom left: density/pressure phase diagram. Bottom right: particle velocity vs. shock velocity for single, double, and triple-shocks. For all plots, green lines represent the principal shock, red is secondary shock, and blue is tertiary shock. Cyan hexagons show the locations of the points indicated in Table 1.

## AUTHOR CONTRIBUTIONS

BM conceived of the project. KDV performed the simulations. KDV, FGC, and BM analyzed the data and wrote the manuscript.

## CONFLICT OF INTEREST

The authors declare no potential conflict of interests.

## ACKNOWLEDGEMENTS

This material is based on work supported by the Department of Energy, National Nuclear Security Administration under award DE-NA0004147. Computational resources at the National Energy Research Scientific Computing Center and the Livermore Computing Center were used.

## DATA AVAILABILITY STATEMENT

The data that support the findings of this study are available from the corresponding author upon reasonable request.

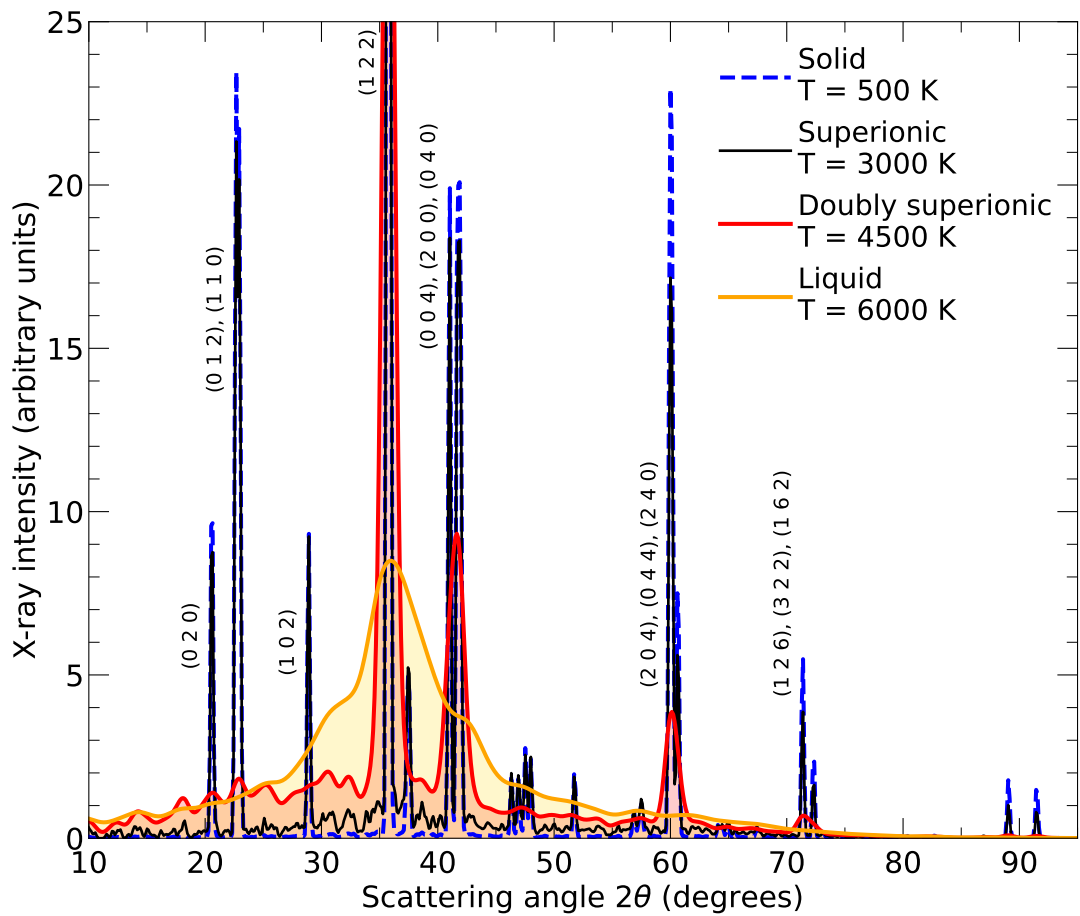

**FIGURE 5** Computed X-ray diffraction spectra of $H_3NO_4$-$P2_12_12_1$ for a density of 6.033 g/cm$^3$. Assuming a wavelength of 1 nm, the spectra were calculated by averaging configurations of MD trajectories for the solid phase at 500 K (dashed blue), the hydrogen superionic phase at 3000 K (black), the doubly superionic (H+N) phase at 4500 K (red), and the liquid phase at 6000 K (orange). The Miller indices are shown for all major peaks. From the solid to the superionic phase, peak intensities decrease only slightly, which is primarily due to thermal motion of the heavy nuclei because hydrogen is a weak X-ray scatterer. During the transition from the hydrogen superionic to doubly superionic phase, the first three peaks disappear because the nitrogen nuclei become disordered. The strongest (1 2 2) reflection persists because the oxygen nuclei remain bound to their lattice sites. When the entire structure melts at 6000 K, this peak marks the maximum of the broad peak of the liquid.